\documentclass[12pt,a4paper]{article}
\usepackage{times}
\usepackage{codams,float,amssymb}
\usepackage[leqno]{amsmath}
\usepackage[dvipdf]{graphicx}
\floatstyle{ruled}
\newfloat{displaycode}{htp}{alg}[section]
\floatname{displaycode}{Eiffel code}
\usepackage[bookmarks,bookmarksopen,
          pdftitle={"An accl not a crcl"},
          pdfauthor={Colm \'O D\'unlaing}]{hyperref}

\newtheorem{definition}[theorem]{Definition}
\newtheorem{lemma}[theorem]{Lemma}

\newtheorem{corollary}[theorem]{Corollary}

\newcommand{\redstar}{{\overset{*}{\rightarrow}}}
\newcommand{\mapstar}{{\overset{*}{\mapsto}}}

\newcommand{\thuecong}{{\overset{*}{\leftrightarrow}}}
\newcommand{\irr}{{\text{\rm irr}}}
\newcommand{\pres}{\,
{\vrule height5pt width.5pt depth0pt
\overset{\vrule height.5pt width10pt depth0pt}
{\vrule height0pt width0pt depth0pt}
\vrule height5pt width.5pt depth0pt}\,}
\newcommand{\presstar}{{\overset{*}{\pres}}}
\numberwithin{equation}{section}

\newcommand{\pos}{{\text{\rm pos}}}
\renewcommand{\neg}{{\text{\rm neg}}}

\newcommand{\be}{\begin{equation*}}
\newcommand{\ee}{\end{equation*}}
\newcommand{\hb}{\hfil\break}

\DeclareGraphicsExtensions{.eps}
\title{An ACCL which is not a CRCL}
\author{Colm \'O D\'unlaing\thanks{e-mail: odunlain@maths.tcd.ie.
Mathematics department website: http://www.maths.tcd.ie.}\\
{\em Mathematics, Trinity College, Dublin 2, Ireland}}

\begin{document}

\maketitle
\begin{abstract}
It is fairly easy to show that every regular set is
an almost-confluent congruential language (ACCL), and it
is known [\ref{dkrw}] that every regular set
is a Church-Rosser congruential language (CRCL).
Whether there exists an ACCL, which is not a CRCL, seems to remain
an open question.  In this note we present one such ACCL.
\end{abstract}

\section{Introduction}

$\Sigma^*$ denotes the set of `strings' over an alphabet
$\Sigma$ --- $\Sigma$ can be any finite set;
strings over $\Sigma$ are finite sequences drawn from $\Sigma$.
$\Sigma^*$ is a  monoid (with identity $\lambda$, the empty string)
under string concatenation.
The length of a string $x$ is denoted $|x|$ ($|\lambda|=0$).
If $x\in \Sigma^*$ and $a\in \Sigma$ then

\be
|x|_a
\ee
is the number of occurrences of $a$ in $x$, so
\be
\sum_{a \in \Sigma} |x|_a = |x|.
\ee

\begin{definition}
\label{def: thue congruence}
A {\em Thue system} over a finite alphabet $\Sigma$
is a set of ordered pairs $(u,w)$ of strings
in $\Sigma^*$.
In this note only finite Thue systems are considered.

If $T$ is a Thue system, then we call the pairs
$(u,w)$ in $T$ its {\em rules}, sometimes written
$\leftrightarrow_T$.

A {\em congruence} on $\Sigma^*$ (or any semigroup) is
an equivalence relation $\equiv$ such that
for all $u,v,x,y \in \Sigma^*$,
\be
x\equiv y\implies uxv \equiv uyv
\ee
The equivalence classes can be multiplied and
thus there is a quotient monoid
\be
\Sigma^* / \equiv .
\ee
If $\equiv$ is a congruence and $x$ a string, we write
\be
[x]_\equiv
\ee
for the congruence class of $x$ modulo $\equiv$.
\end{definition}

\noindent
Given $x,y\in \Sigma^*$, we write
\be
x \leftrightarrow_T y
\ee
if there exist strings $t,u,v,w$, such
that $x = tuv$, $y=twv$, and
either $(u,w)\in T$ or $(w,u)\in T$.

This relation is symmetric, and its
reflexive transitive closure
\be
{\thuecong}_T
\ee
is a congruence on $\Sigma^*$.
The notation for  congruence class is simplified as follows.
\be
[x]_T ~~=\text{\rm (def)}~~ [x]_{{\thuecong}_T}.
\ee

Emphasis is placed on the relative lengths of strings
in rules of $T$.

\noindent
If $x\leftrightarrow_T y$ and in addition $|x|>|y|$, $|x| \geq |y|$, or
$|x|=|y|$, respectively, write
\be
x \to_T y,\quad\text{or}\quad x \mapsto_T y,
\quad\text{or}\quad x {\pres}_T y,
\ee
respectively.

Since the relation $\leftrightarrow_T$ is
symmetric, we can assume that for any
$(u,w)\in T$,
\be
|u| \geq |w|
\ee

\begin{definition}
\label{def: redex}
When $x = tuv \to_T twv = y$, so
$|u| > |w|$, we call $u$ the {\em redex} and $w$ the {\em reduct}.
\end{definition}

\begin{definition}
A Thue system $T$ is, respectively, {\rm (i)} Church-Rosser,
{\rm (ii)} almost confluent, {\rm (iii)} preperfect,
(see {\rm [\ref{bo}]}), if whenever $x {\thuecong}_T y$,

\begin{itemize}
\item [{\rm (i)}] 
there exists a string $z$ such that $x {\redstar}_T z$ and
$y {\redstar}_T z$;
\item [{\rm (ii)}] there exist strings $z_1$ and $z_2$ such that
$x {\redstar}_T z_1$,
$y {\redstar}_T z_2$,
and $z_1 {\presstar}_T z_2$;
\item [{\rm (iii)}]
there exists a string $z$ such that
$x {\mapstar}_T z$ and $y {\mapstar}_T z$.
\end{itemize}
\end{definition}

\begin{definition}
\label{def: irr}
If $T$ is a Church-Rosser Thue system, then for any string
$x$, every string $y$ in $[x]_T$ reduces (modulo $T$) to
the same irreducible string; we call this string
\be
\irr_T(x) .
\ee
\end{definition}

The word problem for Church-Rosser systems is in linear
time, and for the other two kinds it is PSPACE complete;
testing for the Church-Rosser property is tractable;
testing for almost confluence is in PSPACE;
it is undecidable whether a Thue system is preperfect [\ref{bo}].

\begin{definition}
\label{def: congruential language}
A language $L$ is {\em congruential} if there exists
a congruence $\equiv$ and a finite set of strings
\begin{gather*}
 x_1, x_2, \ldots, x_n,\quad\text{ such that}\\
L = [x_1]_\equiv \cup
[x_2]_\equiv \cup \ldots \cup [x_n]_\equiv
\end{gather*}

If the congruence is generated by a Thue system,
i.e., it is ${\thuecong}_T$ for some finite Thue system $T$,
and $T$ is, respectively, Church-Rosser, or almost confluent,
or preperfect, then $L$ is a Church-Rosser, or almost confluent,
or preperfect congruential language: {\em CRCL, ACCL}, or {\em PPCL}.
\end{definition}

An interesting and  old result is that every regular set is
an ACCL.  It can be shown as follows:
if $L$ is a regular set then there exists
a finite monoid $M$ and a homomorphism from
$\Sigma^*$ to $M$ such that $L$ is a union
of $h^{-1}(g)$ for suitable $g$ in $M$.
But this partition 
\be
\{h^{-1}(g): ~ g \in M\}
\ee
can also be realised by a finite almost-confluent
system, namely: let $N$ be the maximal length of minimal
strings in this partition (a $x$ string is minimal
if whenever $x {\thuecong}_T y$, $|x| \leq |y|$).
Then the system
\be
S =\{(x,y):\quad x,y\in \Sigma^*,~
|x| \leq N+1,~ x{\thuecong}_T y,~ y ~ \text{minimal} \}
\ee
is almost confluent and its congruence classes
coincide with the inverse images $h^{-1}(g)$, as required.

A long-standing open problem was
whether every regular set is a CRCL:  it was
settled in the affirmative a few years ago
[\ref{dkrw}].

That left open the unlikely possibility
that every ACCL is a CRCL.  This note shows
the contrary.

The analysis in this paper is simple and direct.
In fact, the problem is not susceptible to more
sophisticated methods.  As noted in [\ref{ods}],
Kolmogorov-complexity-based analyses showing
palindromes not to be Church-Rosser\footnote{Church-Rosser
languages are a much richer class of languages than
Church-Rosser congruential.} also shows them not
to be almost confluent.  Indeed, in [\ref{ods}] we were only able to
show that they are `preperfect languages'.

All Church-Rosser monoids  are $\text{FP}_\infty$ [\ref{squier},\ref{cohen}].
On the other hand, if one inspects
the group furnished by Squier [\ref{squier}],
which is not $\text{FP}_3$, it has an obvious presentation
as a monoid, but the presentation again turns out to be
preperfect rather than almost confluent.

Book's reduction machine [\ref{bo}]
can be used with almost-confluent Thue systems,
from which is follows that ACCLs are linear time recognisable.
The word problem for
an almost confluent Thue system is PSPACE-complete, but
(as is easy to show) if the
system presents a {\em group} then the word problem is linear time.
So there are few complexity-based arguments separating
ACCLs from CRCLs.

\section{An ACCL which is not a CRCL}

We shall introduce an almost confluent Thue system over
a 4-letter alphabet
$\Sigma = \{a,b,c,d\}$, and
an involution
\be
a \mapsto c \mapsto a,\quad
b\mapsto d \mapsto b
\ee
or
\be
\overline{a} = c, \overline{c} = a,
\overline{b} = d, \overline{d} = b.
\ee

\noindent Any string in $\Sigma^*$
can and will be written using $a,b,\overline{a},\overline{b}$.
\begin{definition}
We call $a,b$ {\em positive} and $c,d$ (i.e.,
$\overline{a}, \overline{b}$) {\em negative}.

Given a string $x$ over $a,b,\overline{a},\overline{b}$,

\begin{gather*}
|x|_\pos = |x|_a + |x|_b,\\
|x|_\neg = |x|_{\overline{a}}  + |x|_{\overline{b}},
\end{gather*}
the number of occurrences of positive and negative
letters in $x$.
\end{definition}

Let
\be
h: \Sigma^* \to \IZ
\ee (the additive group of integers) denote the following map:
\be
h(x) ~=~
|x|_\pos - |x|_\neg .
\ee
This is a homomorphism, and
\be
a\mapsto 1, b \mapsto 1,
\overline{a}\mapsto -1, \overline{b}\mapsto -1.
\ee
Let $S$ be the Thue system
\begin{gather*}
a \overline{a} \to \lambda,\quad
\overline{a} a\to \lambda,\quad
a \overline{b} \to \lambda,\quad
\overline{b}a  \to \lambda,\quad
b \overline{a} \to \lambda,\quad
\overline{a} b \to \lambda,\quad
b \overline{b} \to \lambda,\quad
\overline{b} b \to \lambda,\\
a \pres b,\quad b \pres a,
\quad \overline{a} \pres \overline{b},
\quad \overline{b} \pres \overline{a}.
\end{gather*}

\noindent
The map $h$
preserves both sides of each rule in $S$, and
therefore induces a homomorphism
\be
\Sigma^*/ {\thuecong}_S \to \IZ.
\ee

For the rest of this paper, we assume that strings
are written in terms of $a,b,\overline{a}, \overline{b}$.

\begin{definition}
Given a string $x = a_1 a_2 \ldots a_k$, the string
$\tilde{x}$ is defined as
\be
\tilde{x} = \overline{a_k} \, \overline{a_{k-1}} \ldots \overline{a_1}.
\ee
\end{definition}
Clearly $h(x\tilde{x}) = 0$ and
$[x \tilde{x}]_S = [\tilde{x}x]_S = [\lambda]_S$.

\begin{definition}
A string $x$ is {\em mixed} if it contains both positive
($a$ or $b$) and negative ($\overline{a}$ or $\overline{b}$) letters.
Else it is {\em unmixed}.  Unmixed strings can be
empty, positive, or negative, in the obvious sense.
\end{definition}

If $x$ is mixed, then it contains an adjacent pair of
positive and negative letters which can be reduced
(modulo $S$).  Thus mixed strings are reducible.
Unmixed strings are irreducible.

Thus every string $x$ can be reduced to a positive
or negative string.
If $x$ is positive then $h(x) = |x|$.  If $x$
is negative then $h(x) = -|x|$.

\begin{lemma}
If $x$ and $y$ are both positive strings,
or both negative, and $|x| = |y|$, then
$x{\presstar}_S\, y$.\qed
\end{lemma}

\begin{corollary}
$S$ is almost confluent and $h$ induces an
isomorphism of $\Sigma^*/{\thuecong}_S$ with $\IZ$.
\end{corollary}

{\bf Proof.}
Suppose $h(x) = h(y)$.

Reduce $x$ and $y$ (modulo $S$) to irreducible
strings $x'$ and $y'$.  Then $h(x') = h(x) = h(y) = h(y')$,
and $x'$ and $y'$ are unmixed.

If $h(x) = 0$, then $x'=y'=\lambda$.
If $h(x) > 0$, then $x'$ and $y'$ are entirely positive,
$|x'| = |y'|$, and $x' {\presstar}_S y'$.

Similarly if $h(x) < 0$.

We have shown that if $h(x) = h(y)$ then there exist irreducible
strings $x'$ and $y'$ such that $x {\redstar}_S x'$,
$y {\redstar}_S y'$, and $x' {\presstar}_S y'$.

In particular, $x {\thuecong}_S y$.
Conversely, as has been noted, if $x {\thuecong}_S y$ then
$h(x) = h(y)$:
$h$ induces
an isomorphism of $\Sigma^*/{\thuecong}_S$ with its image, $\IZ$.

Finally, if $x {\thuecong}_S y$, then $h(x) = h(y)$,
so there exist strings $x', y'$ so
\be
x {\redstar}_S x' {\presstar}_S y' 
{\overset{*}{\leftarrow}}_S y
\ee
so $S$ is almost confluent.\qed

\begin{definition}
\be
L = [\lambda]_S = h^{-1}(0).
\ee
This is our candidate for a non-CRCL.
\end{definition}

\begin{corollary}
$L$ is an ACCL.\qed
\end{corollary}

\begin{theorem}
\label{thm: main}
$L$ is not a CRCL.
\end{theorem}

We prove this by contradiction. Otherwise there exists
a Church-Rosser Thue system $T$ and a list  of irreducible strings
\be
u_1,\ldots, u_n
\ee
in $L$ such that
\be
\takeanumber
\tag{\thetheorem}
\label{eq: union}
L = [\lambda]_S = [u_1]_T \cup \ldots \cup [u_n]_T
\ee
or equivalently
\be
x \in L \iff \irr_T(x) \in \{u_1, \ldots, u_n\}.
\ee

Associated with $T$ and the strings $u_j$, we define the
following constants:
\begin{definition}
\label{def: QR}
\be
Q = \max_{(\ell,r)\in T} |\ell|
\quad\text{and}\quad
R = \max_{1\leq j\leq n}|u_j|_\neg.
\ee
($Q$ is the maximum length of redexes in $T$.)
\end{definition}

\begin{lemma}
\label{lem: T refines}
If such a Thue system $T$ exists, then $T$
refines $S$ (in the sense
that $x\thuecong_T y \implies x \thuecong_S y$).
\end{lemma}

{\bf Proof.}  It is enough to show that whenever
\be
x \to_T y,
\ee
$[x]_S = [y]_S$. Clearly
\be
x\tilde{x} \to_T y\tilde{x}
\ee
But $x\tilde{x} \in [\lambda]_S$, which is a union of
congruence class modulo $T$, so $y\tilde{x} \in [\lambda]_S$.
Then $[y\tilde{x} x]_S = [\lambda x]_S = [x]_S$.
But $[y\tilde{x}x]_S = [y\lambda]_S = [y]_S$, so
$[x]_S = [y]_S$, as required.\qed

\begin{corollary}
If $x$ is unmixed, then $x$ is irreducible (modulo $T$).
\end{corollary}

{\bf Proof:} $x$ is irreducible (modulo $S$) and $T$ refines
$S$.\qed

\begin{lemma}
\label{lem: reduced by Q}
Suppose that $xy \to_T z$ where $y$ is unmixed (and $|z| \geq Q$).
Then $z$ can be factored as $x'y'$ where $y'$ is  unmixed and $|y'| > |y| - Q$
(\ref{def: QR}).
\end{lemma}

{\bf Proof.} The redex in $xy$ cannot be entirely in
$y$ since $y$ is irreducible.  Therefore the redex
is in $x s$ where $|s| < Q$ (possibly $s =\lambda$).
Setting $xy = xsy'$, $y'$ is a suffix of $z$,
$y'$ is unmixed, and $|y'| > |y| - Q$.\qed

\begin{lemma}
\label{lem: balanced reduction}
Suppose $x \to_T y$. Then $|x|_\pos > |y|_\pos$
and $|x|_\neg  > |y|_\neg$.
\end{lemma}

{\bf Proof}
Since $h(x)=h(y)$, $|x|_\neg-|y|_\neg = |x|_\pos-|y|_\pos$,
so the number of positive and negative letters is reduced
by the same amount, namely, $(|x|-|y|)/2$.\qed

\begin{corollary}
\label{cor: rightmost k}
For any positive integer $k$, if $y$ is positive
of  length $QR+k$ (\ref{def: QR}), then
for $1 \leq i \leq n$,
\be
y\quad\text{\rm and}\quad \irr_T(u_i y)
\ee
agree on their rightmost $k$ letters.
\end{corollary}

{\bf Proof.}
Lemma \ref{lem: reduced by Q} can be extended inductively
so that if $u_i y$ is reduced $t$ times, then
the reduced string agrees with $y$ on their rightmost
$|y|- tQ$ letters.  By Lemma \ref{lem: balanced reduction},
$u_i y$ can be reduced at most $|u_iy|_\neg$ times.
But $|u_iy|_\neg = |u_i|_\neg$ and $|u_i|_\neg \leq R$,
so $y$ and $\irr_T(u_i y)$ agree on their rightmost
$|y|-QR$ letters; and $|y|-QR=k$.\qed

\hb

{\bf Proof of} Theorem \ref{thm: main}.
Let $k=\lceil \log_2 (n+1) \rceil$ and let $x$ be a positive
string of length $QR + k$.  For any positive string
$y$ of the same length as $x$, $x {\presstar}_S y$.

Let $u_i = \irr_T(x\tilde{x})$ (noting that $x\tilde{x}\in L$).
For any positive string $y$ with $|y|=|x|$, $x\thuecong_S y$
so $\tilde{x}x \thuecong_S \tilde{x}y$.  But
$\tilde{x}x \thuecong_S \lambda$, so
$\tilde{x} y \in L$ and $\irr_T(\tilde{x}y) = u_j$ for some $j$.
Therefore $[\tilde{x}y]_T = [u_j]_T$ and
$[x\tilde{x}y]_T = [x u_j]_T$. But $u_i = \irr_T(x\tilde{x})$,
so, for every positive $y$ with $|y|=|x|$,
\be
\takeanumber
\tag{\thetheorem}
\label{eq: mismatch class count}
[u_iy]_T = [x u_j]_T
\ee
for some $j$.
Let $\{y_q\}$ be an enumeration of all positive strings $y$
of length $|x|$ which agree with $x$ on their first $QR$
letters.  There are $2^k$ such strings. By Corollary
\ref{cor: rightmost k}, for each string $y_q$,
\be
y_q \quad\text{\rm and}\quad  \irr_T(u_i y_q)
\ee
agree on their rightmost
$k$ letters.  The irreducible strings belong to
different congruence classes. Therefore there are
at least $2^k$ congruence classes fitting the left-hand
side of equation \ref{eq: mismatch class count}, and there
are at most $n$ classes matching the right-hand side.
Since $2^k>n$, we have  a contradiction: $L$ is not
a CRCL.\qed

\section{Acknowledgement}
The author is grateful to Friedrich Otto
for some corrections and helpful suggestions.

\section{References}
\label{references} 
\begin{enumerate}
\item
\label{bo}
Ronald V.\ Book and Friedrich Otto (1993). {\em String-rewriting
systems.} Springer texts and monographs in computer science.
\item
\label{cohen}
Daniel E.\ Cohen (1997). String rewriting and homology of monoids.
{\em Math.\ Structures in Computer Science \bf 7:3}, 207--240.
\item
\label{dkrw}
Volker Diekert, Manfred Kufleitner, Klaus Reinhardt, and
Tobias  Walter (2012).
Regular languages are Church-Rosser congruential.
{\em Proc.\ 39th.\ ICALP II, Springer LNCS 7392}, 177--188.
\item
\label{ods}
Colm \'O D\'unlaing and Natalie Schluter (2010).
A shorter proof that palindromes are not a Church-Rosser
language, with extensions to almost-confluent and preperfect Thue systems.
{\em Theoretical Computer Science \bf 411}, 677--690.
\item
\label{squier}
Craig C.\ Squier (1987). Word problems and a homological finiteness
condition for monoids. {\em J. Pure and Applied Algebra \bf 49}, 201--217.
\end{enumerate}
\end{document}